\documentclass{article}

\PassOptionsToPackage{numbers}{natbib}

\usepackage[preprint]{neurips_2023}

\usepackage[colorlinks, urlcolor=blue]{hyperref}
\usepackage[utf8]{inputenc} 
\usepackage[T1]{fontenc}    
\usepackage{url}            
\usepackage{booktabs}       
\usepackage{amsfonts}       
\usepackage{nicefrac}       
\usepackage{microtype}      
\usepackage{xcolor}         
\usepackage{amssymb}
\usepackage{bm}
\usepackage{amsmath}
\usepackage{graphicx}
\usepackage{multirow}
\usepackage{pifont}
\usepackage{enumitem}
\usepackage{siunitx}

\def\eqref#1{equation~\ref{#1}}

\def\1{\bm{1}}

\def\vx{{\bm{x}}}
\def\vy{{\bm{y}}}
\def\vz{{\bm{z}}}

\DeclareMathAlphabet{\mathsfit}{\encodingdefault}{\sfdefault}{m}{sl}
\SetMathAlphabet{\mathsfit}{bold}{\encodingdefault}{\sfdefault}{bx}{n}

\usepackage{cancel}

\newcommand\blfootnote[1]{%
  \begingroup
  \renewcommand\thefootnote{}\footnote{#1}%
  \addtocounter{footnote}{-1}%
  \endgroup
}

\title{MusicLDM: Enhancing Novelty in Text-to-Music Generation Using Beat-Synchronous Mixup Strategies}

\author{%
  \quad\textbf{Ke Chen}$^{1*}$
  \quad\textbf{Yusong Wu}$^{2*}$
  \quad\textbf{Haohe Liu}$^{3*}$
  \quad\textbf{Marianna Nezhurina}$^{4}$ \\
  \quad\textbf{Taylor Berg-Kirkpatrick}$^{1}$
  \quad\textbf{Shlomo Dubnov}$^{1}$\\ \\ 
  $^{1}$University of California San Diego \\ 
  $^{2}$Mila, Quebec Artificial Intelligence Institute, Université de Montréal \\ 
  $^{3}$Centre for Vision Speech and Signal Processing, University of Surrey \\ 
  $^{4}$LAION \\
}

\begin{document}

\maketitle

\begin{abstract}
Diffusion models have shown promising results in cross-modal generation tasks, including text-to-image and text-to-audio generation.\blfootnote{*The first three authors have equal contribution. Contact: Ke Chen (knutchen@ucsd.edu), Yusong Wu (wu.yusong@mila.quebec), and Haohe Liu (haohe.liu@surrey.ac.uk)}. 
However, generating music, as a special type of audio, presents unique challenges due to limited availability of music data and sensitive issues related to copyright and plagiarism. In this paper, to tackle these challenges, we first construct a state-of-the-art text-to-music model, MusicLDM, that adapts Stable Diffusion and AudioLDM architectures to the music domain. We achieve this by retraining the contrastive language-audio pretraining model (CLAP) and the Hifi-GAN vocoder, as components of MusicLDM, on a collection of music data samples. Then, to address the limitations of training data and to avoid plagiarism, we leverage a beat tracking model and propose two different mixup strategies for data augmentation: beat-synchronous audio mixup and beat-synchronous latent mixup, which recombine training audio directly or via a latent embeddings space, respectively. Such mixup strategies encourage the model to interpolate between musical training samples and generate new music within the convex hull of the training data, making the generated music more diverse while still staying faithful to the corresponding style. In addition to popular evaluation metrics, we design several new evaluation metrics based on CLAP score to demonstrate that our proposed MusicLDM and beat-synchronous mixup strategies improve both the quality and novelty of generated music, as well as the correspondence between input text and generated music.
\end{abstract}

\section{Introduction}

Text-guided generation tasks have gained increasing attention in recent years and have been applied to various modailties, including text-to-image, text-to-video, and text-to-audio generation. Text-to-image generation has been used to create both realistic and stylized images based on textual descriptions, which can be useful in various scenarios including graphic design. Text-to-audio generation is a relatively new, but rapidly growing area, where the goal is to generate audio pieces, such as sound events, sound effects, and music, based on textual descriptions. Diffusion models have shown superior performance in these types of cross-modal generation tasks, including systems like DALLE-2 \cite{dalle-2-ramesh2022hierarchical} and Stable Diffusion \cite{stable-diffusion-rombach2022high} for text-to-image; and AudioGen \cite{kreuk2022audiogen}, AudioLDM \cite{audioldm-liu2023audioldm}, and Make-an-Audio \cite{make-an-audio-huang2023make} for text-to-audio.

As a special type of audio generation, \textit{text-to-music} generation has many practical applications \cite{briot2020deep, ml-algo-creative-musical-tool-fiebrink2016machine}. For instance, musicians could use text-to-music generation to quickly build samples based on specific themes or moods and speed up their creative process. Amateur music lovers could leverage generated pieces to learn and practice for the purpose of musical education. 
However, text-to-music generation presents several specific challenges. One of the main concerns is the limited availability of text-music parallel training data \cite{agostinelli2023musiclm}. Compared to other modalities such as text-to-image, there are relatively few text-music pairs available, making it difficult to train a high-quality conditional model. Large and diverse training sets may be particularly imperative for music generation, which involves many nuanced musical concepts, including melody, harmony, rhythm, and timbre. Further, the effectiveness of diffusion models trained on more modest training sets has not been fully explored. Finally, a related concern in text-to-music generation is the risk of plagiarism or  lack of novelty in generated outputs \cite{agostinelli2023musiclm}. Music is often protected by copyright laws, and generating new music that sounds too similar to existing music can lead to legal issues. Therefore, it is important to develop text-to-music models that can generate novel and diverse music while avoiding plagiarism, even when trained on relatively small training datasets. 

In this paper, we focus on both these challenges: we develop a new model and training strategy for learning to generate novel text-conditioned musical audio from limited parallel training data. Currently, since there is no open-source model for text-to-music generation, we first construct a state-of-the-art text-to-music generation model, MusicLDM, which adapts the Stable Diffusion \cite{stable-diffusion-rombach2022high} and AudioLDM \cite{audioldm-liu2023audioldm} architectures to the music domain.
Next, to address the limited availability of training data and to encourage novel generations, we adapt an idea from past work in other modalities: mixup \cite{zhang2017mixup}, which has been applied to computer vision and audio retrieval tasks, augments training data by recombining existing training points through linear interpolation. This type of augmentation encourages models to interpolate between training data rather than simply memorizing individual training examples, and thus may be useful in addressing data limitations and plagiarism in music generation. However, for music \textit{generation}, the naive application of mixup is problematic. Simply combining waveforms from two distinct musical pieces leads unnatural and ill-formed music: tempos and beats (as well as other musical elements) are unlikely to match. Thus, we propose two novel mixup strategies, specifically designed for music generation: beat-synchronous audio mixup (BAM) and beat-synchronous latent mixup (BLM), which first analyze and beat-align training samples before interpolating between audio samples directly or encoding and then interpolating in a latent space, respectively. 

We design new metrics that leverage a pretrained text and audio encoder (CLAP) to test for plagiarism and novelty in text-to-music generation. In experiments, we find that our new beat-synchronous mixup augmentation strategies, by encouraging the model to generate new music within the convex hull of the training data, substantially reduce the amount of copying in generated outputs. Further, our new model, MusicLDM, in combination with mixup, achieves better overall musical audio quality as well as better correspondence between output audio and input text. In both automatic evaluations and human listening tests, MusicLDM outperforms state-of-the-art models at the task of text-to-music generation while only being trained on 9K text-music sample pairs.

    
    
Music samples and qualitative results are available at \href{https://musicldm.github.io}{musicldm.github.io}. Code and models are available at \href{https://github.com/RetroCirce/MusicLDM/}{https://github.com/RetroCirce/MusicLDM/}.

\section{Related Work}

\subsection{Text-to-Audio Generation}
Text-to-audio generation (TTA) \cite{audioldm-liu2023audioldm, kreuk2022audiogen, yang2023diffsound} is a type of generative task that involves creating audio content from textual input. In years past, text-to-speech (TTS) \cite{ren2020fastspeech, tan2022naturalspeech} achieved far better performance than other types of audio generation. However, with the introduction of diffusion models, superior performance in various generation tasks became more feasible. Recent work has focused on text-guided generation in general audio, with models such as Diffsound \cite{yang2023diffsound}, AudioGen \cite{kreuk2022audiogen}, AudioLDM \cite{audioldm-liu2023audioldm}, and Make-an-Audio \cite{make-an-audio-huang2023make} showing impressive results. In the domain of music, text-to-music models include the retrieval-based MuBERT \cite{MubertAI}, language-model-based MusicLM \cite{agostinelli2023musiclm}, diffusion-based Riffusion \cite{Forsgren_Martiros_2022} and Noise2Music \cite{huang2023noise2music}. However, a common issue with most recent text-to-audio/music models is the lack of open-source training code. Additionally, music models often requires large amounts of privately-owned music data that are inaccessible to the public, which makes it difficult for researchers to reproduce and build upon their work. 

Among all these models, AudioLDM is based on open-source Stable Diffusion \cite{stable-diffusion-rombach2022high}, CLAP \cite{clap-wu2023large}, and HiFi-GAN \cite{kong2020hifi} architectures. Therefore, we base our text-to-music generation model on the AudioLDM architecture, to create MusicLDM for our experiments.

\subsection{Plagiarism on Diffusion Models}
Diffusion models have been shown to be highly effective at generating high-quality and diverse samples for text-to-image tasks. However, a potential issue with these models is the risk of plagiarism \cite{somepalli2022diffusion, carlini2023extracting}, or the generation novelty. As stated by \cite{somepalli2022diffusion}, diffusion models are capable of memorizing and combining different image objects from training images to create replicas, which can lead to highly similar or even identical samples to the training data. \cite{carlini2023extracting} explores different methods that could extract the training data with a generate-and-filter pipeline, showing that new advances in privacy-preserving training of diffusion models are required. Such issues are especially concerning in the domain of music, where copyright laws are heavily enforced and violations can result in significant legal and financial consequences. Therefore, there is a need to develop strategies to mitigate the risk of plagiarism in text-to-music generation using diffusion models.

\vspace{-0.2cm}
\subsection{Mixup on Data Augmentation}
Mixup \cite{zhang2017mixup} is a widely used data augmentation technique that has shown remarkable success in improving model generalization and mitigating overfitting. The basic principle of mixup is to linearly combine pairs of training samples to effectively construct new samples that lie on the line connecting the original samples in the feature space, encouraging the model to learn a more continuous and robust decision boundary. In this paper, we explore the mixup technique in the context of text-to-music generation based on latent diffusion models. Different from the mixup in other modalities, music mixup involves a delicate balance of musical elements to prevent the mixed music from being chaotic noise. Moreover, in diffusion models, mixup also can refer to the combination of latent features, rather than music signals. We propose two mixup strategies tailored for  music latent diffusion models and explore their potential benefits for data augmentation and generation performance.

\begin{figure}
    \centering
    \includegraphics[width=\textwidth]{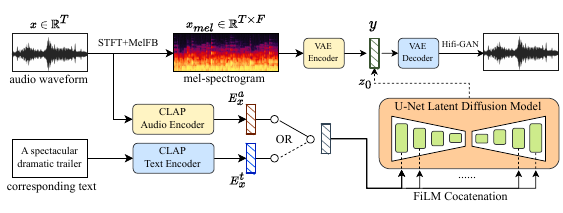}
    \vspace{-0.8cm}
    \caption{The architecture of MusicLDM, which contains a contrastive language-audio pretraining (CLAP) model, an audio latent diffusion model with VAE, and a Hifi-GAN nerual vocoder.}
    \label{fig:musicldm}
    \vspace{-0.3cm}
\end{figure}

\vspace{-0.2cm}
\section{Methodology}

\subsection{MusicLDM}

As illustrated in Figure \ref{fig:musicldm}, MusicLDM has similar architecture as AudioLDM: a contrastive language-audio pretraining (CLAP) model \cite{clap-wu2023large}, an audio latent diffusion model \cite{audioldm-liu2023audioldm} with a pretrained variational auto-encoder (VAE) \cite{kingma2013auto}, and a Hifi-GAN neural vocoder \cite{kong2020hifi}. 

Formally, given an audio waveform $\vx \in \mathbb{R}^{T}$ its corresponding text, where $T$ is the length of samples, we feed the data into three modules: 
\begin{enumerate}[leftmargin=*]
    \item We pass $\vx$ through the audio encoder \cite{chen2022hts} of CLAP $f_{audio}(\cdot)$, to obtain the semantic audio embedding $\bm{E}_{x}^{a} \in \mathbb{R}^D$, where $D$ is the embedding dimension.
    
    \item We pass the text of $x$ through the text encoder \cite{liu2019roberta} of CLAP $f_{text}(\cdot)$, to to obtain the semantic text embedding $\bm{E}^{t}_{x} \in \mathbb{R}^D$.

    \item We transform $\vx$ into in the mel-spectrogram $\vx_{mel} \in \mathbb{R}^{T \times F}$. Then we pass $\vx_{mel}$ into the VAE encoder, to obtain an audio latent representation $\vy \in \mathbb{R}^{C \times \frac{T}{P} \times \frac{F}{P}}$, where $T$ is the mel-spectrogram frame size, $F$ is the number of mel bins, $C$ is the latent channel size of VAE, and $P$ is the downsampling rate of VAE. The VAE is pretrained to learn to encoder and decode the mel-spectrogram of music data. 
\end{enumerate}

In MusicLDM, the latent diffusion model has a UNet architecture where each encoder or decoder block is composed of a ResNet layer \cite{he2016deep} and a spatial transformer layer \cite{stable-diffusion-rombach2022high}. 
During the training, the semantic embedding of the input audio $\bm{E}_x$ is concatenated with the latent feature of each UNet encoder and decoder block by the FiLM mechanism \cite{perez2018film}. The output of the diffusion model is the estimated noise $\bm{\epsilon}_\theta(\vz_n,n,\bm{E}_x)$ from $n$ to $(n-1)$ time step in the reverse process, where $\theta$ is the parameter group of the diffusion model, and $\vz_n$ is the $n$-step feature generated by the forward process. We adopt the training objective \cite{DDPM, DiffWave} as the mean square error (MSE) loss function:
\begin{align}
    L_n(\theta)=\mathbb{E}_{\vz_0,\bm{\epsilon},n} ||\bm{\epsilon} - \bm{\epsilon}_\theta(\vz_n,n,\bm{E}_x)||_2^2 
\end{align}
where $\vz_0=\vy$ is the audio latent representation from VAE (i.e., the groundtruth), and $\bm{\epsilon}$ is the target noise for training. More details regarding the training and the architecture of the latent diffusion model can be referred in Appendix A.

For MusicLDM, we make two changes from the original AudioLDM to enhance its performance on text-to-music generation. 
First, since the original contrastive language-audio pretraining (CLAP) model is pretrained on text-audio pair datasets dominated by sound events, sound effects and natural sounds, we retrained the CLAP on text-music pair datasets (details in Appendix B) to improve its understanding of music data and corresponding texts. We also retrained the Hifi-GAN vocoder on music data to ensure high-quality transforms from mel-spectrograms to music waveforms. 
Second, in the original AudioLDM, the model is only fed with audio embeddings as the condition during the training process, i.e., $\bm{E}_x = \bm{E}_x^a$; and it is fed with text embeddings to perform the text-to-audio generation, i.e., $\bm{E}_x = \bm{E}_x^t$. This approach leverages the alignment of text and audio embeddings inside CLAP to train the latent diffusion model with more audio data without texts. However, this audio-to-audio training $\bm{\epsilon}_\theta(\vz_n,n,\bm{E}_x^a)$ is essentially an approximation of the text-to-audio generation $\bm{\epsilon}_\theta(\vz_n,n,\bm{E}_x^t)$. Although CLAP is trained to learn joint embeddings for text and audio, it does not explicitly enforce the embeddings to be distributed similarly in the latent space, which can make it challenging for the model to generate coherent text-to-audio outputs solely with audio-to-audio training. This problem becomes more severe when the available text-music pair data is limited. Moreover, relying solely on audio embeddings ignores the available text data, which means that we are not leveraging the full potential of our dataset. Consequently, generating accurate and realistic text-to-audio generations may not be effective.

To further investigate this task, we introduce two additional training approaches for comparison:
\begin{enumerate}[leftmargin=*]
    \item Train the MusicLDM directly using the text embedding as the condition, i.e., $\bm{\epsilon}_\theta(\vz_n,n,\bm{E}_x^t)$
    \item Train the MusicLDM using the audio embedding as the condition, then finetune it with text embedding, i.e.,$\bm{\epsilon}_\theta(\vz_n,n,\bm{E}_x^a) \rightarrow \bm{\epsilon}_\theta(\vz_n,n,\bm{E}_x^t)$
\end{enumerate}
The first approach follows the original target of text-to-audio, serving as a comparison with audio-to-audio training. The second approach is proposed as an improvement on audio-to-audio generation.
as we shift the condition distribution from the audio embedding back to the text embedding during the training of the diffusion model. 
In section \ref{sec:music-quality-experiment}, we compared the above two approaches with the original audio-to-audio training approaches to determine the best approach for generating high-quality and highly correlated text-to-music outputs.

\subsection{Beat-Synchronous Mixup}\label{sec:mixup}
\begin{figure}
    \centering
    \includegraphics[width=\textwidth]{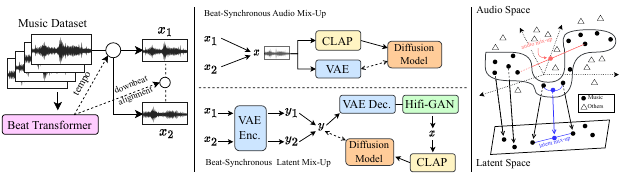}
    \caption{Mixup strategies. Left: tempo grouping and downbeat alignment via Beat Transformer. Middle: BAM and BLM mixup strategies. Right: How BAM and BLM are applied in the feature space of audio signals and audio latent variables.}
    \label{fig:mixup}
    \vspace{-0.25cm}
\end{figure}

As shown in Figure \ref{fig:mixup}, we propose two mixup strategies to augment the data during the MusicLDM training: Beat-Synchronous Audio Mixup (BAM) and Beat-Synchronous Latent Mixup (BLM). 

\paragraph{Beat-tracking via Beat Transformer}
Musical compositions are made up of several elements, such as chord progressions, timbre, and beats. Of these, beats play a crucial role in determining the musical structure and alignment. In most audio retrieval tasks, mixup is a popular technique that involves randomly mixing pairs of audio data to augment the training data. However, when mixing two music samples that have different tempos (beats per minute), the mixture can be chaotic and unappealing.
To avoid this, we use a state-of-the-art beat tracking model, Beat Transformer \cite{zhao2022beat}, to extract the tempo and downbeat map of each music track, as shown in the left of Figure \ref{fig:mixup}. We categorize each music track into different tempo groups and during training, we only mixed tracks within the same tempo group to ensure the tracks were in similar tempos. Furthermore, we align the tracks by comparing their downbeat maps and selecting a certain downbeat to serve as the starting position for the mixup track. This preprocessing approach allows us to better select the music data available for mixup, resulting in mixup tracks that are neatly ordered in terms of tempo and downbeats.

\paragraph{Beat-Synchronous Audio Mixup}
As depicted in the upper part of Figure \ref{fig:mixup}, once we select two aligned music tracks $\vx_1$ and $\vx_2$, we mix them by randomly selecting a mixing ratio from the beta distribution $\lambda \sim \mathcal{B}(5,5)$, as:
\begin{align}
    \vx = \lambda \vx_1 + (1-\lambda) \vx_2
\end{align}
We then use the mixed data $\vx$ to obtain the CLAP embedding $\bm{E}_x$ and the audio latent variable $\vy$. We train the latent diffusion model using the standard pipeline. This beat-synchronous audio mixup strategy is referred to as BAM.

\paragraph{Beat-Synchronous Latent Mixup}
As depicted in the lower part of Figure \ref{fig:mixup}, in the latent diffusion model, the mixup process can also be applied on the latent variables, referred as beat-synchronous latent mixup (BLM). After selecting two aligned music tracks $\vx_1$ and $\vx_2$, we feed them into the VAE encoder to obtain the latent variables $\vy_1$ and $\vy_2$. We then apply the mixup operation to the latent variables:
\begin{align}
    \vy = \lambda \vy_1 + (1-\lambda) \vy_2
\end{align}

In contrast to BAM, BLM applies the mixup operation to the latent space of audio, where we cannot ensure that the mixture of the latent variables corresponds to the actual mixture of the music features in the appearance. Therefore, we first generate a mixed mel-spectrogram $\vx_{mel}$ by feeding the mixed latent variable $\vy$ into the VAE decoder. Then, we feed $\vx_{mel}$ to the Hifi-GAN vocoder to obtain the mixed audio $\vx$ as the input music. With $\vx$ and $\vy$, we follow the pipeline to train the MusicLDM.

\paragraph{What are BAM and BLM doing?}
As shown in the right of Figure \ref{fig:mixup}, we demonstrate the interpolation between the feature space of audio when using BAM and BLM. In the feature space of audio signals, the "$\bullet$" represents the feature point of music data, while the "$\triangle$" denotes the feature point of other audio signals, such as natural sound, audio activity, and noise. During the pretraining process of VAE, a latent space is constructed for encoding and decoding the music data. The VAE aims to learn the distribution of the latent variables that can best represent the original data and transform the original feature space into a lower-dimensional manifold. This manifold is designed to capture the underlying structure of the music data. Therefore, any feature point within this manifold is considered to be a valid representation of music.


BAM and BLM are concerned with augmenting the data at different levels of feature space. As shown in right of Figure \ref{fig:mixup}, BAM linearly combines two points in audio space to form a new point on the red line. BLM, represented by the blue line, performs a similar operation, but result in a new point in the VAE-transformed latent space, which will be decoded back onto the music manifold of audio space.

Both BAM and BLM offer unique advantages and disadvantages. BAM applies mixup in the original feature space, resulting in a smooth interpolation between feature points. However, BAM cannot ensure a reasonable music sample that lies within the music manifold. This issue is more problematic using the simple audio mixup strategy without tempo and downbeat alignments. BLM, conversely, augments within the music manifold, fostering robust and diverse latent representations. However, BLM is computationally more expensive as it requires computing the latent feature back to audio via VAE decoder and Hifi-GAN. 
Furthurmore, when the ill-defined or collapsed latent exists in VAE, BLM may be out of effectiveness.

Both BAM and BLM are effective data augmentation techniques that encourage the model to learn a more continuous and robust decision boundary on the audio feature space, or implicitly from the latent space to the audio space, which can improve the model's generalization performance and mitigate overfitting. In the context of text-to-music generation,  these mixup strategies can have a potential to mitigate the limitations of data size and help avoid plagiarism issues. By introducing small variations through mixup, the model can touch a more rich space of music data and generate music samples that are correlated to the texts but show differences to the original training data. In Section \ref{sec:music-quality-experiment}, we evaluated whether these strategies mitigate the data limitation and plagiarism issues.

\section{Experiments}
In this section, we conducted four experiments on our proposed methods. First, we retrained a new CLAP model to provide the condition embedding for MusicLDM. Second, we trained MusicLDM with different mixup strategies and compared them with available baselines. Third, we evaluated MusicLDM in terms of text-music relevance, novelty and plagiarism risk via metrics based on CLAP scores. Finally, we conducted a subjective listening test to give an additional evaluation.

\subsection{Experimental Setup}

\paragraph{Dataset}

The original CLAP model trained on mostly acoustic events and sound effect datasets. In this work, we trained a CLAP model on music datasets in addition to its original training data, allowing it to better understand the relation between music and textual descriptions. The new CLAP model is trained on dataset of 2.8 Million text-audio pairs, in an approximate total duration of \num{20000} hours. Compared to previous CLAP model, the newly trained CLAP model performs well in zero-shot classification for both acoustic event and music. Please refer further details on training and performance of the new CLAP in Appendix B. For MusicLDM, 
we used the Audiostock dataset for training, along with VAE and Hifi-GAN. Specifically, the Audiostock dataset contains \num{9000} music tracks for training and \num{1000} tracks for testing. The total duration is \num{455.6} hours. It provides a correct textual description of each music track.
Although CLAP is trained on more text-music data pairs, the large number of them are pseudo-captions composed primarily of non-specific metadata, such as author, song title, and album information (e.g., [\texttt{a song by author A from the album B}]). These captions do not align with our specific objective of text-to-music generation.

\paragraph{Hyperparameters and Training Details}

We trained all MusicLDM modules with music clips of 10.24 seconds at \SI{16}{\kilo\hertz} sampling rate. In both the VAE and diffusion model, music clips are represented as mel-spectrograms with $T=1024$ frames and $F=128$ mel-bins. Unlike AudioLDM, MusicLDM's VAE utilizes a downsampling rate of $P=8$ and a latent dimension of $C=16$. The architecture of MusicLDM's latent diffusion model follows that of AudioLDM. The training process of MusicLDM aligns with AudioLDM's approach. For additional hyperparameters and training details, refer to Appendix A.

\begin{table}[t]
\caption{The evaluation of generation quality among MusicLDMs and baselines. AA-Train. and TA-Train. refer to the audio-audio training scheme and the text-audio traning scheme.}
\label{tab:ttm-result}
\centering
\resizebox{\textwidth}{!}{
\begin{tabular}{lcc|cccc}
\toprule
Model & AA-Train. & TA-Train. & FD$_{pann}$ $\downarrow$ & FD$_{vgg}$ $\downarrow$ & Inception Score $\uparrow$ & KL Div. $\downarrow$ \\ \midrule
Riffusion \cite{Forsgren_Martiros_2022}  & \ding{55} & \ding{51}  & 68.95 & 10.77 & 1.34 & 5.00 \\
MuBERT \cite{MubertAI}  & --- & --- & 31.70 & 19.04 & 1.51 & 4.69 \\
AudioLDM & \ding{51}  & \ding{55} & 38.92 & 3.08 & 1.67 & 3.65 \\ \midrule
MusicLDM & \ding{51}   & \ding{55} & 26.67 & 2.40 & \textbf{1.81} & 3.80 \\
MusicLDM (Only TA-Training)  & \ding{55} & \ding{51} & 32.40 & 2.51 & 1.49 & 3.96 \\
MusicLDM w/. mixup & \ding{51}  & \ding{55} & 30.15 & 2.84 & 1.51 & 3.74 \\
MusicLDM w/. BAM   & \ding{51}  & \ding{55} & 28.54 & \textbf{2.26} & 1.56 & 3.50 \\
MusicLDM w/. BLM & \ding{51}   & \ding{55} & \textbf{24.95} & 2.31 & 1.79 & \textbf{3.40} \\ \midrule
MusicLDM w/. Text-Finetune  & \ding{51}  & \ding{51}  & 27.81  & 1.75 & 1.76  & 3.60  \\
MusicLDM w/. BAM \& Text-Finetune  & \ding{51}  & \ding{51}  & 28.22 & 1.81 & 1.61 & 3.61 \\
MusicLDM w/. BLM \& Text-Finetune  & \ding{51}  & \ding{51}  & \textbf{26.34} & \textbf{1.68} & \textbf{1.82} & \textbf{3.47} \\ \bottomrule
\end{tabular}
}
\vspace{-0.5cm}
\end{table}

\subsection{MusicLDM Performance}\label{sec:music-quality-experiment}

\subsubsection{Generation Quality}
Following evaluation techniques used in past work on audio generation \cite{audioldm-liu2023audioldm}, we use frechet distance (FD), inception score (IS), and kullback-leibler (KL) divergence to evaluate the quality of generated musical audio outputs. Frechet distance evaluates the audio quality by using an audio embedding model to measure the similarity between the embedding space of generations and that of targets. In this paper, we use two standard audio embedding models: VGGish \cite{hershey2017cnn} and PANN \cite{pann}. The resulting distances we denote as $FD_{vgg}$ and $FD_{pann}$, respectively. Inception score measures the diversity and the quality of the full set of audio outputs, while KL divergence is measured on individual pairs of generated and groundtruth audio samples and averaged. We use the \texttt{audioldm\_eval} library\footnote{https://github.com/haoheliu/audioldm\_eval} to evaluate all the metrics mentioned above, comparing the groundtruth audio from the Audiostock 1000-track test set with the 1000 tracks of music generated by each system based on the corresponding textual descriptions.


Table \ref{tab:ttm-result} presents the FD, IS, and KL results for our models in comparison with baseline models. In the first section, we utilized textual descriptions from the test set, sending them to the offical APIs of Riffusion and MuBERT to generate corresponding results. Both Riffusion and MuBERT were unable to achieve results comparable to the remaining models.
Upon reviewing the generated music from the two systems, we found that the sub-optimal performance of Riffusion resulted from poor music generation quality, with many samples either inaudible or outside the desired musical range. MuBERT, while generating high-quality pieces from real music sample libraries, fell short in replicating the distribution of Audiostock dataset. Due to the unavailability of their detailed architectures, training scripts, and data, Riffusion and MuBERT's evaluations offered only partial comparisons.


We also retrained the original AudioLDM model on the Audiostock dataset, comparing it to MusicLDM variants. The distinction between AudioLDM and MusicLDM lies in the different CLAP models used for condition embeddings. Our comparison revealed that MusicLDM outperforms AudioLDM in terms of $FD_{pann}$, $FD_{vgg}$, and IS. This underscores the efficacy of the novel CLAP model pretrained for music, providing more suitable music embeddings as conditioning information.


Comparing MusicLDM's performance with audio-to-audio training ($\bm{\epsilon}_\theta(\vz_n,n,\bm{E}x^a)$) and text-to-audio training ($\bm{\epsilon}_\theta(\vz_n,n,\bm{E}x^t)$), denoted by ``MusicLDM (Only TA-Training)'', we note inferior results in the latter approach. This may be suggests that a gap exists between distribution of text embedding and audio embedding, making it challenging for the diffusion model to generate high-quality audio solely from text embedding. In contrast, CLAP's audio embedding may leak low-level audio information to the diffusion model during initial training stages, hurting the model's ability to generalize to text embedding inputs. This hypothesis is further supported by the results of MusicLDM with combined audio-to-audio training and text-to-audio fine-tuning. We observe a significant decrease in $FD{vgg}$ with small changes in $FD{pann}$ and IS, indicating a substantial improvement in music generation quality, driven by leveraging both audio and text embeddings during training. The former facilitates good audio reconstruction during early training, while the latter shifts the distribution from audio to text to align with the eventual test-time task of text-to-music generation.


Last, we compared MusicLDM with different mixup strategies, namely simple mixup \cite{zhang2017mixup}, BAM, and BLM. The comparison reveals the negative impact of the simple mixup on all metrics. This degradation in generated sample quality, characterized by instrumental interference and noise, is attributed to the simple mixup's inability to guarantee musicality in the mix. Similar observations are evident in the BAM results, indicated by a drop in $FD_{pann}$ and IS. However, BAM's tempo and downbeat alignment, along with the original mixup benefits, counterbalance this defect to a certain extent, enhancing the model's generalization ability and improving certain metrics. BLM, as a latent space mixing method, aligns with our hypothesis in Section \ref{sec:mixup} that latent space mixup yield audio closely resembling music. This technique allows us to largely bypass the potential confusion issues tied to audio mixing, thus capitalizing on mixup's ability to drive generalization and prevent copying via data augmentation. Furthermore, incorporating text-finetuning results in a comprehensive improvement of music generation quality, solidifying BLM as the most effective strategy.

\begin{table}[t]
\caption{The objective metrics to measure the relevance and novelty (plagiarism). And the subjective listening test to evaluate the quality, relevance, and musicality. }
\label{tab:os_results}
\resizebox{\textwidth}{!}{
\begin{tabular}{l|ccc|ccc}
\toprule
\multirow{3}{*}{Model} & \multicolumn{3}{c|}{Objective Metrics} & \multicolumn{3}{c}{\multirow{2}{*}{Subjective Listening Test}} \\ \cmidrule{2-4}
 & \multicolumn{1}{c|}{Relevance} & \multicolumn{2}{c|}{Novelty and Plagiarism Risk} & \multicolumn{3}{c}{} \\ \cmidrule{2-7} 
 & \multicolumn{1}{c|}{Text-Audio Similarity$\uparrow$} & $SIM_{AA}@90\downarrow$ & $SIM_{AA}@95\downarrow$ & Quality$\uparrow$ & Relevance$\uparrow$ & Musicality$\uparrow$ \\ \midrule
Test Set (Ref.) & \multicolumn{1}{c|}{0.325} & --- & --- & --- & --- & --- \\
Retrieval Max (Ref.) & \multicolumn{1}{c|}{0.423} & --- & --- & --- & --- & --- \\ \midrule 
MuBERT & \multicolumn{1}{c|}{0.131} & 0.107 & 0 & 2.02 & 1.50 & \textbf{2.33} \\
MusicLDM (original) & \multicolumn{1}{c|}{\textbf{0.281}} & 0.430 & 0.047 & 1.98 & 2.17 & 2.19 \\
MusicLDM w/. mixup & \multicolumn{1}{c|}{0.234} & \textbf{0.391} & 0.028 & --- & --- & --- \\
MusicLDM w/. BAM & \multicolumn{1}{c|}{0.266} & 0.402 & 0.027 & 2.04 & 2.21 & 2.01 \\
MusicLDM w/. BLM & \multicolumn{1}{c|}{0.268} & 0.401 & \textbf{0.020} & \textbf{2.13} & \textbf{2.31} & 2.07 \\ \bottomrule
\end{tabular}}

\end{table}

\begin{figure}[t]
    \centering
    \includegraphics[width=\textwidth]{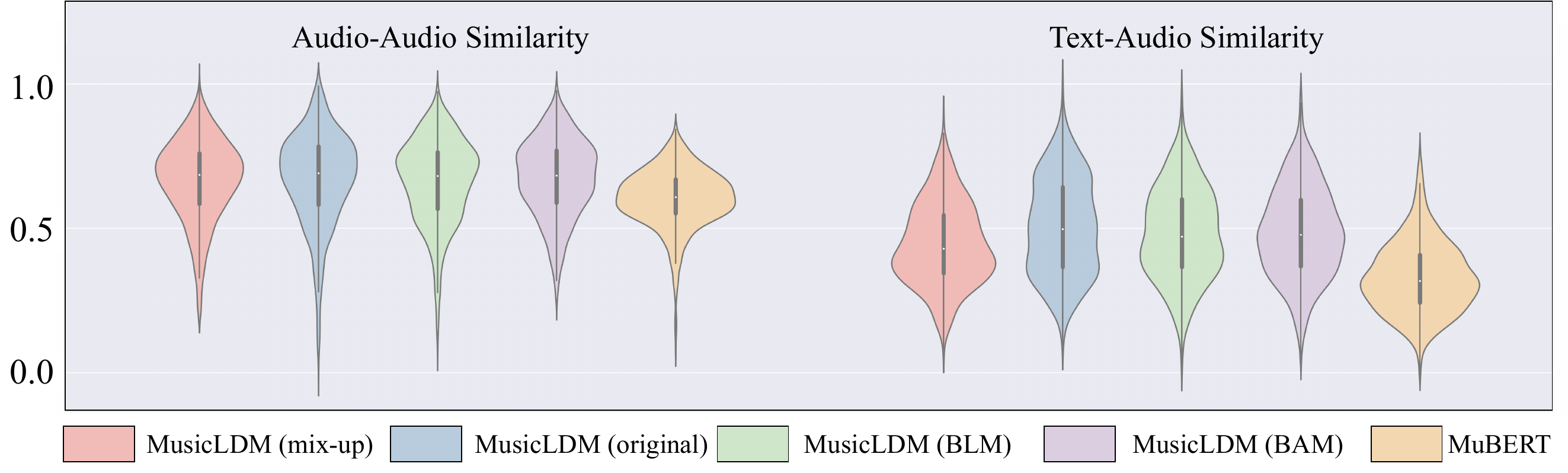}
    \caption{The violin plot of the audio-audio similarity, and the text-to-audio similarity.}
    \label{fig:clapscore_violin}
    \vspace{-0.5cm}
\end{figure}

\subsubsection{Text-Audio Relevance, Novelty and Plagiarism} \label{sec:sim-aa}


We proposed two metric groups, \textbf{text-audio similarity} and \textbf{nearest-neighbor audio similarity ratio} to assess text-audio relevance, novelty, and plagiarism risk in various models.


First, text-audio similarity measures the relevance between the text and the audio. It is defined as the dot product between the groundtruth text embedding $\bm{E}_{gd}^t$ from the test set and the audio embedding $\bm{E}^a$ from music generated by models, i.e., $\bm{E}_{gd}^t \cdot \bm{E}^a$. The embeddings from both text and audio are normalized in CLAP model, thus the dot product computes the cosine similarity between text and audio embeddings.

Second, we would also like to measure the extent to which models are directly copying samples from the training set. We verify this by first computing the dot products between the audio embedding of each generated music output to all audio embeddings from the Audiostock training set and returning the maximum -- i.e., the similarity of the nearest-neighbor in the training set. Then, we compute the fraction of generated outputs whose nearest-neighbors are above a threshold similarity. We refer this as the nearest-neighbor audio similarity ratio, providing $SIM_{AA}@90$ where the threshold is 0.9, and $SIM_{AA}@95$ with 0.95. 
The lower this fraction, the lower the risk of plagiarism -- i.e., fewer samples have very close training neighbors. In the Appendix D, we show  pairs of examples with both high and low similarity scores to give further intuition for this metric.

As shown in the left and middle column of Table \ref{tab:os_results}, we present the average text-audio similarity and nearest-neighbor audio similarity ratios for two thresholds on the 1000 tracks in the Audiostock test set and the generated music from MuBERT and different variants of MusicLDM. We also provide two reference points for text-audio similarity: ``Test Set'' and ``Retrieval Max''. Specifically, ``Test Set'' refers to computing the cosine similarity between the groudtruth text embedding and the groudtruth audio embedding. And ``Retrieval Max'' refers to first computing the cosine similarities between each text embedding from the test set to all audio embeddings from the training set, then picking the highest score as the score of this text, and taking the average over all text scores. 

We can observe that the original MusicLDM without mixup achieves the highest text-audio relevance with an average score of 0.281, but 
also the highest (worst) nearest-neighbor audio similarity ratio. MusicLDM with the simple mixup strategy achieves the lowest $SIM_{AA}@90$ ratio while sacrificing a lot in the relevance of the generation. The MusicLDM with BAM and BLM achieve a balance between the audio similarity ratios and the text-to-audio similarity. In combination with the quality evaluation results in Table \ref{tab:ttm-result}, we can conclude that all mixup strategies are effective as a data augmentation techniques to improve generalization of the model to generate more novel music. However simple mixup degrades the generation quality, which affects the relevance score between audio and text, and also thus makes it less similar to the tracks in the training set. BAM and BLM apply the tempo and downbeat filtering on the music pairs to mix, allowing the model to maintain superior generation quality (Table \ref{tab:ttm-result}) and text-audio relevance (Table \ref{tab:os_results}), while still utilizing the benefit brought by the mixup technique to make the generation more novel (less plagiarism). Among the objective metrics, BLM is the best mixup strategy in terms of quality, relevance and novelty of the generated audio. This indicates mixing in the latent space is more efficient than mixing directly in audio space, perhaps because the latent embedding approach implicitly projects the mixture to the learned manifold of well-formed music. 
We show the detailed distribution of these metrics over 1000 generated tracks in Figure \ref{fig:clapscore_violin}, where, for example, audio-audio similarity denotes the individual scores used to calculate the average $SIM_{AA}$. We find that the original MusicLDM without mixup has more samples with high training similarity than other models, which further reflects that it is more prone to copying. 

\subsection{Subjective Listening Test}
As shown in the right of Table \ref{tab:os_results}, we conduct the subjective listening test on four models, namely MuBERT, the original MusicLDM, and that with BAM or BLM strategy, to further evaluate the actual hearing experience of the generated music. We do not include the simple mixup MusicLDM because its generation is at a low quality while we avoid confusing subjects with too many models in the same time. We invite 15 subjects to listen to 6 groups of the generations randomly selected from the test set. Each of group contains four generations from four models and the corresponding text description. The subjects are required to rate the music in terms of quality, relevance, and musicality (detailed in Appendix E).
We observe that the samples of MusicLDM with BAM or BLM mixup strategy achieve a better relevance and quality than those of MuBERT and the original MusicLDM, this strengths our above analysis. The MuBERT samples achieve the best Musicality, because its generation is combined from the real music samples. Combined with the objective metrics, beat-synchronous latent mixup stands to be the most effectiveness method for enhancing the text-to-music generation in terms of quality, text-music relevance and novelty (i.e., reducing the risk of plagiarism).  

\vspace{-0.25cm}
\section{Limitations}
\vspace{-0.25cm}
In this section we outline the recognized limitations of our study, serving as a roadmap for future improvements.
Firstly, MusicLDM is trained on the music data in a sampling rate of \SI{16}{\kilo\hertz}, while most standard music productions are \SI{44.1}{\kilo\hertz}. This constraint, tied to the Hifi-GAN vocoder's subpar performance at high sampling rates, impedes practical text-to-music application and necessitates further improvements.
Secondly, resource constraints such as limited real text-music data and GPU processing power prevent us from scaling up MusicLDM's training. We are unable to determine if mix-up strategies could yield similar trends as observed with the Audiostock dataset. This issue exists in the image generation task as well. 
Lastly, while we recognize beat information as crucial for music alignment, there is scope for exploring other synchronization techniques like key signature and instrument alignment. We also intend to investigate the application of different audio space filters to select suitable music pairs for mixing.

\section{Conclusion}
In this paper, we introduce MusicLDM, a text-to-music generation model that incorporates CLAP, VAE, Hifi-GAN, and latent diffusion models. We enhance MusicLDM by proposing two efficient mixup strategies: beat-synchronous audio mixup (BAM) and beat-synchronous latent mixup (BLM), integrated into its training process. We conduct comprehensive evaluations on different variants of MusicLDM using objective and subjective metrics, assessing quality, text-music relevance, and novelty. The experimental results demonstrate the effectiveness of BLM as a standout mixup strategy for text-to-music generation. 

\section{Acknowledgments}
We would like to thank the Institute for Research and Coordination in Acoustics and Music (IRCAM) and Project REACH: Raising Co-creativity in Cyber-Human Musicianship for supporting this project. This project has received funding from the European Research Council (ERC REACH) under the European Union's Horizon 2020 research and innovation programme (Grant Agreement \#883313). We would like to thank the support of computation infrastructure from LAION.

\bibliographystyle{plain}  
\bibliography{refs}

\clearpage
\appendix
\thispagestyle{empty}
\section{MusicLDM Details}\label{sec:appendix-model-detail}

\paragraph{Hyperparameters}
For audio signal processing, we use the sampling rate of \SI{16}{\kilo\hertz} to convert all music samples for the training of MusicLDM. Each input data is a chunk of 10.24 seconds randomly selected from the dataset, i.e., $L=163840$. We use the hop size \num{160}, the window size \num{1024}, the filter length \num{1024}, and the number of mel-bins \num{128} to compute the short-time Fourier transform (STFT) and mel-spectrograms. As the result, the input mel-spectrogram has the time frame $T=1024$ and the mel-bins $F=128$.  

We adopt a convolutional VAE as the latent audio representation model, consisting of a 4-block downsampling encoder and a 4-block upsampling decoder. The downsampling and upsampling rate $P=8$ and the latent dimension $C=16$, i.e., the bottleneck latent variable $y$ has a shape of $(C \times \frac{T}{P} \times \frac{F}{P})= (16 \times 128 \times 16)$. For the latent diffusion model, we refer the UNet latent diffusion model\footnote{https://huggingface.co/spaces/multimodalart/latentdiffusion}. It contains 4 encoder blocks, 1 bottleneck block, and 4 decoder blocks. Each block contains 2 residual CNN layers \cite{cnn} and 1 spatial transformer layer \cite{transformer}. The channel dimensions of encoder blocks are \num{128}, \num{256}, \num{384}, and \num{640} and reversed in decoder blocks. For Hifi-GAN, we adopt its official repository\footnote{https://github.com/jik876/hifi-gan} along with the configuration\footnote{https://github.com/jik876/hifi-gan/blob/master/config\_v1.json}. We change the number of mel-bins to \num{128} to fit the processing of MusicLDM. 

\paragraph{Implementation and Training Details}
For the training of VAE, we use the Adam optimizer with a learning rate of \num[scientific-notation=true]{0.0000045} with a batch size of \num{24}. We apply the mel-spectrogram loss, adversarial loss, and a Gaussian constraint loss as the training object of VAE. For the training of Hifi-GAN, we use the batch size of 96 and the AdamW optimizer with $\beta_1=0.8$, $\beta_2=0.99$ at the learning rate of \num[scientific-notation=true]{0.0002}. For the training of MusicLDM, we use the batch size of 24 and the AdamW optimizer with the basic learning of \num[scientific-notation=true]{0.00003}. In the forward process, we use 1000-step of a linear noise schedule from $\beta_1=0.0015$ to $\beta_{1000}=0.0195$. In the sampling process, we use the DDIM \cite{song2020denoising} sampler with 200 steps. We adopt the classifier-free guidance \cite{ho2022classifier} with a guidance scale $w=2.0$. When applying the mixup strategy, we use the mixup rate $p=0.5$. The CLAP model is trained on 24 A100 GPUs. The VAE and HifiGAN model are trained on 4 A60 GPUs. Last, the latent diffusion model is trained on single NVIDIA A40. All models are converged at the end of the training. 

\paragraph{Implementation of Comparison Model}

For generating from Riffusion and MuBERT, we use the official API of Riffusion\footnote{https://huggingface.co/riffusion/riffusion-model-v1} and MuBERT\footnote{https://github.com/MubertAI/Mubert-Text-to-Music}.

\section{CLAP Details}\label{sec:clap}

\begin{table}[t]
\caption{Comparison of zero-shot classification performance of the CLAP used in this work with previous audio-language contrastive learning models.}
\label{tab:clap-perf}
\centering
\begin{tabular}{@{}lcccc@{}}
\toprule
                       & ESC-50 & US8K & VGGSound & GTZAN \\ \midrule
Wav2CLIP \cite{wu2022wav2clip}              & 41.4   & 40.4 & 10.0     & -     \\
audioCLIP \cite{guzhov2022audioclip}              & 68.6   & 68.8 & -        & -     \\
CLAP (Elizalde et al. \cite{elizalde2023clap}) & 82.6   & 73.2 & -        & 25.2  \\
CLAP (Wu et al. \cite{clap-wu2023large})       & 91.0   & 77.0 & 46.2     & 71.0  \\
CLAP (ours on music data)      & 90.1   & 80.6 & 46.6     & 71.0  \\ \bottomrule
\end{tabular}
\end{table}

\subsection{Hyperparameters}

For model hyperparameters, we refer to the official repository\footnote{https://github.com/LAION-AI/CLAP} to conduct the training process of CLAP. The audio encoder of CLAP is HTS-AT-base \cite{chen2022hts} and the text encoder is RoBERTa-base \cite{liu2019roberta}. The HTS-AT-base model has an embedding dimension of 128, and a window size of 8. The HTS-AT-base model has 4 groups of swin-transformer blocks, each group has depth of $[2,2,12,2]$ and number of head in $[4,8,16,32]$. The RoBERTa-base consists of a transformer model with 12 layers, 8 heads, and a inner width of 512. The audio embedding and the text embedding have the dimension size $D=512$. 

\subsection{Training Details}

For the training of CLAP, we use the batch size of \num{2304} and the Adam \cite{kingma2014adam} optimizer with $\beta_1=0.99$, $\beta_2=0.9$ with a warm-up \cite{goyal2017accurate} and cosine learning rate decay at a basic learning rate of \num[scientific-notation=true]{0.0001}.

\subsection{Zero-shot Classification Performance}
We follow previous works on audio-language contrastive learning \cite{goyal2017accurate} to evaluate the performance of CLAP on the zero-shot audio classification tasks, namely on the benchmark datasets of ESC-50 \cite{esc50}, UrbanSound 8K \cite{us8k}, and VGGSound \cite{vggsound}. To demonstrate that the retrained CLAP involves more understandings of music data, we further add a music genre classification benchmark dataset GTZAN \cite{gtzan} into the evalution. As shown in Table \ref{tab:clap-perf}, our retained CLAP achieves best performance acoustic event classification in UrbanSound 8K and VGGSound dataset, while still maintaining comparable performance in ESC-50 dataset and on par performance in GTZAN music classification dataset. Although the performance on GTZAN music dataset is not improved, the extra data used for training CLAP might result in a better representation space which is beneficial for text-to-music generation model.

\begin{figure}
    \centering
    \includegraphics[width=\textwidth]{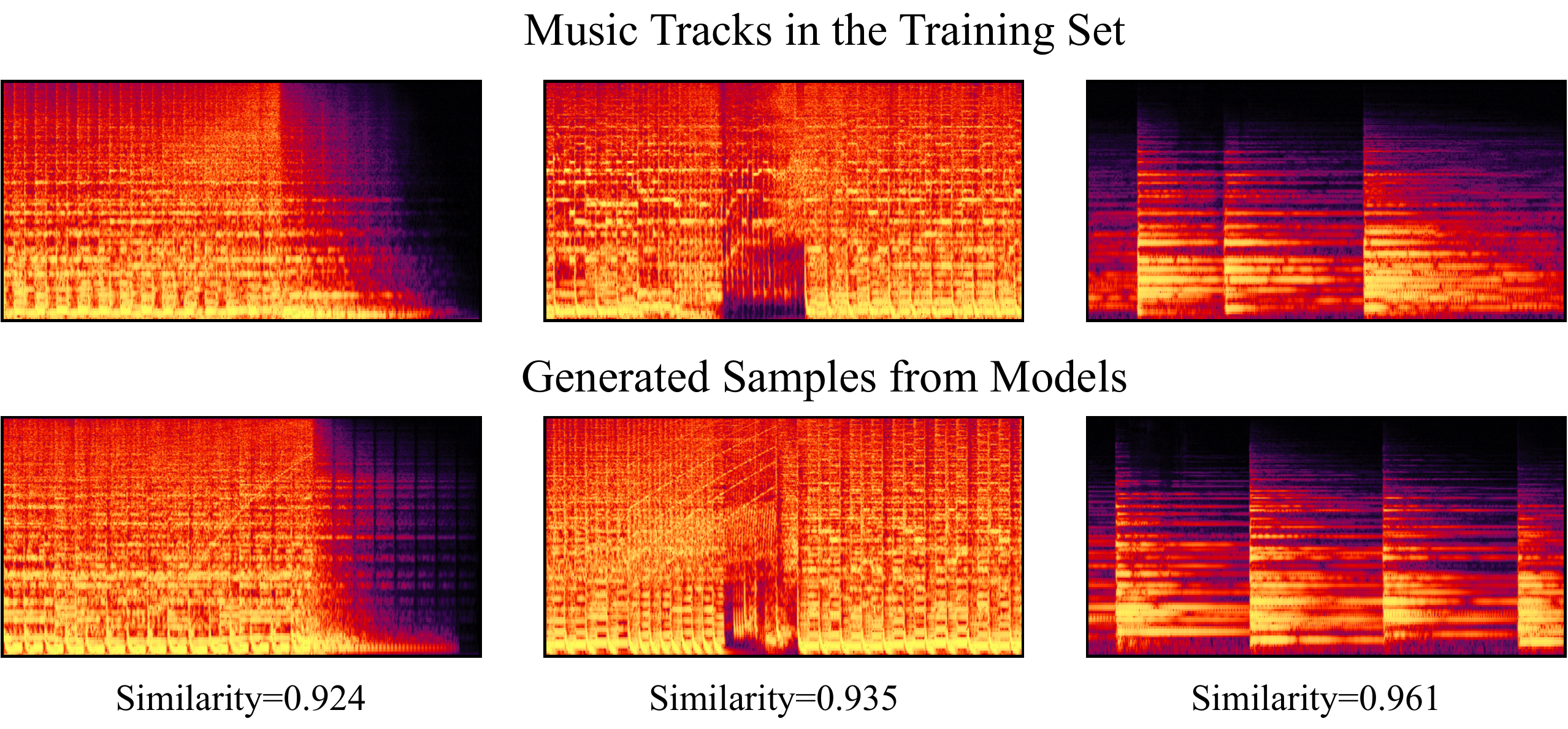}
    \caption{The spectrograms of music pairs indicated by high cosine similiarity score of CLAP audio embeddings.}
    \label{fig:sim_audio}
\end{figure}

\begin{figure}
    \centering
    \includegraphics[width=\textwidth]{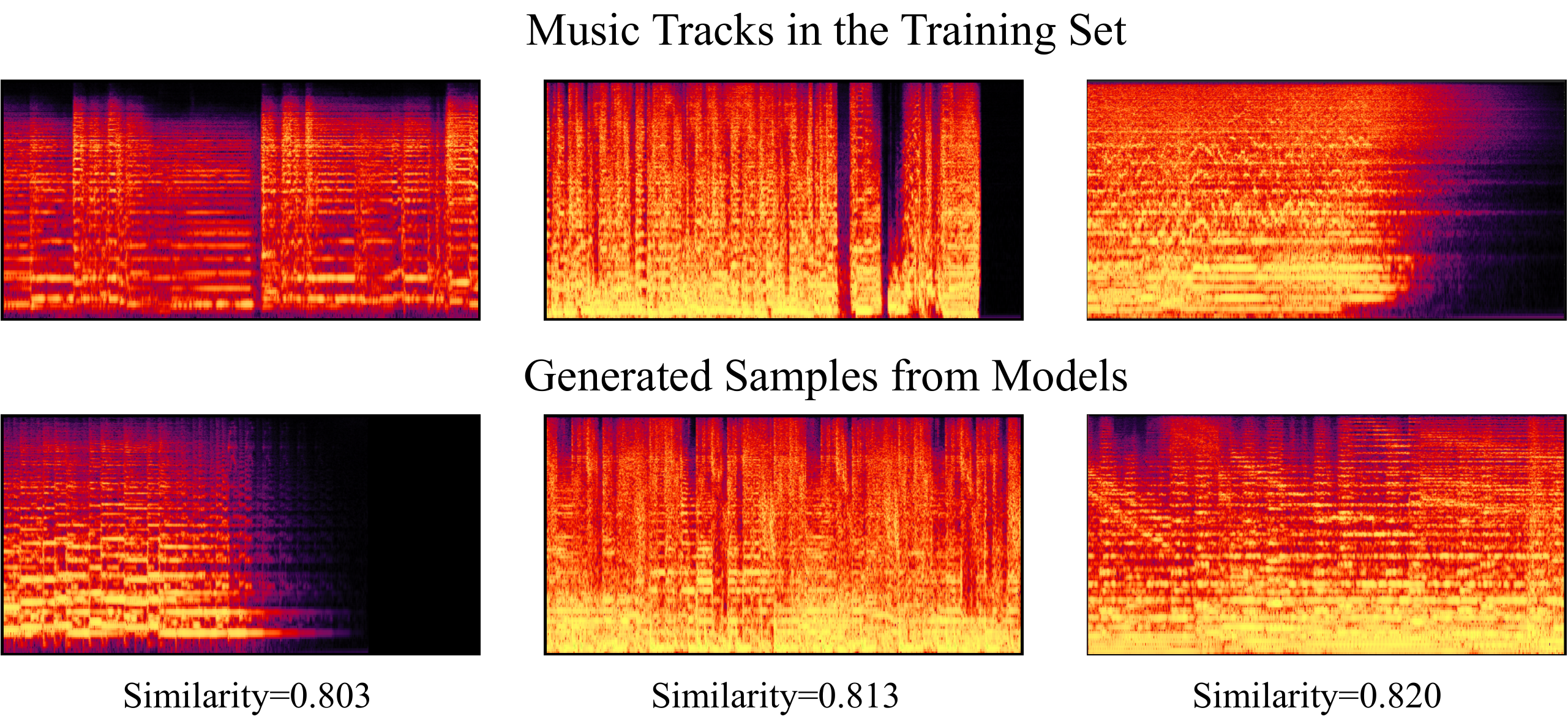}
    \caption{The spectrograms of music pairs indicated by low cosine similarity score of CLAP audio embeddings.}
    \label{fig:disim_audio}
\end{figure}

\section{Nearest-Neighbor Audio Similarity Samples} \label{sec:nn-asim}
As mentioned in section \ref{sec:sim-aa}, we introduced the computation of the nearest-neighbor audio similarity ratio by comparing the cosine similarity between generated music and music tracks in the training set of Audiostock.

In this section, we provide visualizations of the similarity between the generated music and the training music using spectrograms, showcasing how well the cosine similarity between CLAP audio embeddings captures this similarity.

As shown in Figure \ref{fig:sim_audio} and Figure \ref{fig:disim_audio}, display both three examples of music pairs with high and low similarity. To achieve this, we divided the training music tracks into 10-second segments and determined the most similar segment to the generated music track (i.e., the query track).

For instances with high similarity, the cosine similarity of CLAP audio embeddings reveals highly similar structural patterns, indicating a close resemblance in the music arrangements. Conversely, low CLAP cosine similarity indicates significant differences between the spectrograms of the generated music and the training music. This demonstrates the effectiveness of CLAP embeddings in assessing the similarity between music tracks and serving as a means to detect novelty and potential instances of plagiarism in the generated samples.

\section{Subjective Listening Test} \label{sec:subjective}

The subjective listening test was conducted in an online survey format to gather feedback and insights on the text-to-music generation using MusicLDMs and MuBERT. The generation of Riffusion was not included due to its lower quality and relevance compared to the standard. The test had an estimated duration of approximately 10 minutes.

At the beginning of the test, participants were asked to provide their age range and music background as metadata. Subsequently, participants were randomly assigned six groups of generated songs. Each group consisted of four songs generated from MusicLDM, MusicLDM with BAM, MusicLDM with BLM, and MuBERT, all based on the same textual description. The order of the songs within each group was shuffled to eliminate positional bias during rating. Participants were required to rate each song based on three metrics:

\begin{itemize}
    \item Relevance: Determine how well the song matches the given music description. Rate the song based on how closely it aligns with the provided description. 

    \item Quality: Assess the overall quality of the music. Consider factors such as clarity, absence of noise, and general audio quality. Rate the song based on these criteria.

    \item Musicality: Evaluate the musical attributes of the song, including rhythm, melodies, and textures. Rate the song based on its overall musical appeal.
\end{itemize}

Each song in the subjective listening test had a duration of approximately 10 seconds and included a fade-in and fade-out to mitigate bias from the song's beginning and ending sections. The rating scale used for evaluating the songs was designed such that a higher score indicates better quality. Participants were asked to rate each song based on the provided metrics, taking into account the song's overall quality, relevance to the given text, and personal preference on its musicality.

\section{Broader Impact}

The development and implementation of MusicLDM, or generally a text-to-music generation model offers potential benefits and also raises concerns that must be addressed responsibly.

\paragraph{Positive Impacts}

\begin{itemize}
    \item Promoting Creativity: This model can serve as a tool to augment human creativity. Artists, composers, and music amateurs can use it to transfer their textual ideas into music, broadening the realm of artistic exploration and making music creation more accessible and convenient.

    \item Cultural Preservation and Evolution: The model provides a unique platform to archive, digitize, and even evolve cultural musical expressions. Textual descriptions of traditional and folk music can be transformed into the actual music, thereby helping to preserve heritage while simultaneously allowing for creative adaptations. Literature, such as poetry, can be interpreted as music to explore more relations between different types of cultural expression forms. 

    \item Education and Research: In academia, this model can be used as a pedagogical tool in music education. It can aid in understanding the complex relationship between music and linguistic structures, enriching interdisciplinary research in musicology, linguistics, and artificial intelligence.

    \item Entertainment Industry Innovation: The entertainment industry could use this model to generate soundtracks for movies, games, and other media based on scripts. This could potentially revolutionize the way music is produced for media, reducing time and costs.
\end{itemize}

\paragraph{Negative Impacts}

\begin{itemize}
    \item Artistic Job Displacement: While this model can augment human creativity, it may also lead to job losses in the music industry if widely adopted for composing and production. The model could potentially replace human composers in certain contexts, particularly in industries such as film and gaming that require a significant amount of background music.

    \item Copyright Issues: In this paper, one of the targets is to mitigate the copyright issues and plagiarism. The generated music could unintentionally resemble existing works, raising complex copyright infringement issues. It is crucial to implement measures to ensure that the model does not violate intellectual property rights. 

    \item Ethical Misuse: The model could be misused to create music promoting hate speech, misinformation, or other harmful content if the input text has such characteristics. Thus, it is essential to develop safeguards to mitigate the risk of misuse.

    \item Cultural Appropriation and Homogenization: While the model can help preserve music, there is a risk of homogenizing unique cultural musical styles or misappropriating them without proper credit or context.

\end{itemize}

The design and application of this model should be carried out responsibly, considering the potential ethical, social, and economic consequences. Balancing its many benefits with its potential downsides will require the collective effort of developers, users, policy makers, and society at large.

\end{document}